\documentstyle[12pt]{article}
\begin{document}

\def\ti{\tilde}

\def\vt{\tilde{v}}

\def\al{\alpha}
\def\alt{\tilde{\alpha}}
\def\la{\langle}
\def\ra{\rangle}

\def\th{\theta}
\def\dpm{\partial_{\pm}}

\def\be{\begin{equation}}
\def\ee{\end{equation}}

\def\si{\sigma}
\def\cg{{\cal G}}
\def\cm{{\cal M}}
\def\cgt{\ti{\cal G}}
\def\ck{{\cal K}}
\def\ckt{\ti{\cal K}}
\def\cd{{\cal D}}
\def\ce{{\cal E}}

\def\e{\varepsilon}
\def\b{\beta}

\begin{titlepage}
\vskip 1cm
\begin{flushright}
CERN-TH/95-248\\
hep-th/9509095
\end{flushright}
\vskip 1cm
\begin{center}
{\Large\bf Poisson-Lie T-duality} \\
\vskip 2cm
{\bf C. Klim\v c\'\i k\footnote{e-mail: klimcik@surya20.cern.ch}
} \\
\vskip 0.3cm
  {\it Theory Division CERN, CH-1211 Geneva 23,
Switzerland} \\

\end{center}
\vskip 2cm
\begin{abstract}

A description of dual non-Abelian duality is given, based on
the notion of the Drinfeld double. The presentation basically
follows the original paper \cite{KS2}, written in collaboration with
P. \v Severa, but here the emphasis is put  on the algebraic rather
than the geometric aspect of the construction and  a concrete
example of the Borelian double is worked out in detail.
\vskip 1cm
\begin{center}
 {\it Lecture given at}\\
 {\it Trieste Conference on S-Duality
and Mirror Symmetry, June 1995.}
\end{center}
\end{abstract}
\vskip 0.5cm
\noindent CERN-TH/95-248

\noindent September 1995
\end{titlepage}

\section{\bf Introduction}

Duality symmetries  are
expected to
play an important role in string theory as
tools for disentangling its full symmetry structure. They are specific to
string physics and their study led to important insights in understanding
the geometry of space-time from the string point of view.
At present, much is known about the Abelian T-duality
 \cite{Busch} and also the mirror symmetry \cite{BRG}.
In the contribution  \cite{OQ}
(see also previous works \cite{FJ,FT}) there appeared an interesting
possibility of non-Abelian generalization of the standard Abelian T-duality.
For a $\sigma$-model possessing a global non-Abelian isometry with respect
to the group $G$, a non-Abelian dual was found which, however, turned out
to lack the isometry that  would make it possible to perform the duality
transformation back to the original model. In the series of subsequent
 investigations
\cite{GR} (for a review see
\cite{AAL2,GPR})
other relevant insights were
obtained, but the understanding of the sense in which both models were
`dual' to each other still was basically missing. The latter problem
was solved in \cite{KS2} where it was shown that the models are indeed
dual in the sense of the `Poisson-Lie' T-duality.

In the present contribution, based on the  paper with P. \v Severa
\cite{KS2}, I will describe the properties of the Poisson-Lie T-duality;
in particular I will advocate the idea that the relevant  structure
underlying the (non-Abelian) T-duality is the Drinfeld double.
In what follows, I shall construct mutually dual pairs of $\si$-models
for an arbitrary Drinfeld double; the duality transformation will simply
exchange the roles of the two groups forming the double. Then I shall
consider the criteria under which a $\si$-model has its Poisson-Lie dual
and, finally, I will present  explicit forms of the dual pair of models
for the Borelian double.

\section{\bf Manin triples and T-duality}

In this section I will describe the construction of a dual pair
of $\si$-models which are equivalent in the sense of having
the same field equations and the symplectic structure of their
phase spaces. They are dual in the new `Poisson-Lie' sense
which generalizes\footnote{Recently, a connection was also found  in
\cite{Tyu2} between
non-Abelian axial-vector duality \cite{KO} and Poisson-Lie T-duality.} the
Abelian T-duality \cite{Busch} and  the non-Abelian T-duality between
a group and its Lie algebra \cite{OQ,FJ,FT,GR}.
 For the description of the Poisson-Lie duality I shall need the
crucial concept
of the Drinfeld double,  which is  any Lie group $D$ such that
its Lie algebra
$\cd$ can be decomposed into a pair of maximally isotropic subalgebras with
respect to a non-degenerate invariant bilinear form on $\cd$ \cite{D}.
 The isotropic subspace of $\cd$ is such that the value of the form
on two arbitrary
 vectors belonging to the subspace vanishes; maximally isotropic means
that the subspace cannot be enlarged while preserving the property of isotropy.
  Any such decomposition of the double into the pair of maximally
isotropic subalgebras
$\cg + \cgt=\cd $ I shall refer to as the Manin triple.

 Denote
$M({\cal D})$ the set of the Manin triples corresponding to a given
Drinfeld double. It plays the role of the modular space of $\si$-models
mutually connected by Poisson-Lie  T-duality transformation.
(In the Abelian case the Drinfeld double is $D=U(1)^{2d}$ and its modular
space is nothing but $M(\cd_{Abel})=O(d,d,Z)$). Let us also remark that in
general $M(\cd)$ has always at least two points, i.e. $\cg+\cgt =\cd $ and
$\cgt+\cg =\cd $.

A classification of various T-dualities is essentially given by the
following types
of  underlying Drinfeld doubles:

\noindent i) Abelian doubles, which correspond to the standard
Abelian T-duality \cite{Busch};

\noindent ii) semi-Abelian doubles
 (there exists the decomposition
$\cg+\cgt=\cd$ such that
$\cgt$ is Abelian)\footnote{The semi-Abelian double can  always be viewed
as the cotangent bundle of the group manifold $G$ with the group structure
given by the semi-direct product of $G$ and its coalgebra $\cgt$ on which
$G$ acts by the coadjoint action.}, which correspond to the standard
non-Abelian T-duality
between a $G$-isometric $\si$-model with a $G$-target and a non-isometric
$\si$-model with the target $\cgt$ viewed as the Abelian group \cite{OQ,FJ,FT};

\noindent iii) non-Abelian doubles (all the others). They correspond to
the non-trivial
Poisson-Lie T-duality described in \cite{KS2} where none of the models
from the dual pair is isometric with respect to the action of the group, which
naturally acts on its target.

In what follows, I give a unique description of T-duality valid for every
Drinfeld double; for the special cases of the types i) and ii) I recover
the  results known previously.
 I will call the group $G$ with the algebra $\cg$ the duality group
and the group $\ti G$ with the algebra $\cgt$ the coduality group, if
the decomposition of the double is written as $\cd =\cg+\cgt$.
We shall see in a while that for every such decomposition of
the  double there exist
  $\si$-models such that the duality group $G$ acts freely on their
targets and this action is Poisson-Lie symmetric with respect to
the coduality group $\ti G$; the notion of the
Poisson-Lie symmetry, I will explain in section 3.
 Every such $\si$-model has its dual
counterpart  for which the
role of the duality and coduality groups is interchanged. The dual
model has the same field equations as the original one, of course,
in appropriate variables. Moreover, the duality map is a symplectomorphism
between the phase spaces of both models \cite{KS2}.

For the sake of clarity, I shall first consider the `atomic duality' case
in which the
duality group acts on the resulting
$\si$-model target not only freely but also transitively,
which means that the target itself can be identified with the group
manifold. If the duality group does not act transitively on the
$\si$-model target, the free action means that the target is a principal
$G$-bundle; I will refer to this situation as to the `Buscher's duality' case
and  describe it afterwards.
In the Abelian case with $U(1)$ duality group
and another $U(1)$ coduality group, the atomic duality means
that the target space is a one-dimensional circle; in other words, this
is the standard $R\to 1/R$ duality. In the case of the semi-Abelian double,
the most standard example \cite{OQ,FJ,FT} of
 atomic duality is provided by the principal chiral
model on a simple group $G$
\be L(g) =Tr(g^{-1}\partial_- g) (g^{-1}\partial_+ g)\ee
and its dual
\be \ti L(\chi)=\ti E_{ab}(\chi)\partial_- \chi^a \partial_+ \chi^b. \ee
Here $\chi^a$ are coordinates of the elements of coalgebra $\cgt$ and
the matrix $E_{ab}(\chi)$ is given by
\be (\ti E_{\chi}^{-1})^{ab}=\delta^{ab}+\chi^k c_k^{ab},\ee
 $c_k^{ab}$ being the structure constants of the Lie algebra $\cg$ of $G$.
The coduality group in this case is the Abelian group with the same  number
of generators as the dimension of $\cg$. These results may be derived
by choosing the Drinfeld doubles of types i) and ii), using the
following method
valid for a general double:

Consider  an $n$-dimensional linear subspace $\ce^+$ of the
($2n$-dimensional)
Lie algebra
$\cd$ and its orthogonal complement $\ce^-$ such that
$\ce^+ + \ce^-$ span the whole algebra $\cd $. I shall show that those
data determine a dual pair of the $\si$-models with the targets being
groups $G$ and $\ti G$  \cite{KS2}. Indeed, consider
the following field equations for the mapping $l(\xi^+,\xi^-)$ from the
world-sheet of string
into the Drinfeld double $D$ considered as a group:
\be \langle \dpm l l^{-1},\ce^{\pm}\rangle=0.\ee
Here obviously the bracket $\la .,.\ra$ means the invariant bilinear
form on the double.
According to Drinfeld, there exists
the unique decomposition (at least in the vicinity of the unit
element of $D$) of any arbitrary element of $D$ as the product of elements
from $G$ and $\ti G$, i.e.
\be l(\xi^+,\xi^-)=g(\xi^+,\xi^-)\ti h(\xi^+,\xi^-).\ee
Inserting this ansatz into Eq. (1) we obtain:
\be \langle g^{-1}\dpm g +\dpm \ti h \ti h^{-1},g^{-1}\ce^{\pm}g\rangle=0.\ee
It is convenient to introduce mutually dual bases
$T^i$ and $\ti T_i$ in both algebras $\cg$ and
$\cgt$ respectively, which means that
\be \langle T^i,\ti T_j\rangle =\delta^i_j.\ee
Then it is also convenient to write
\be g^{-1}\ce^+ g ={\rm Span}(T^i + E^{ij}(g)\ti T_j), i=1,\dots,n,\ee
\be g^{-1}\ce^- g ={\rm Span}(T^i - E^{ji}(g)\ti T_j), i=1,\dots,n.\ee
By inserting  (8) and (9) into (6), it follows that
\be -(\partial_+\ti h \ti h^{-1})^i=
E^{ij}(g)(g^{-1}\partial_+ g)_j\equiv A_+^i(g),\ee
\be -(\partial_-\ti h \ti h^{-1})^i=
-E^{ji}(g)(g^{-1}\partial_- g)_j\equiv A_-^i(g).\ee
Now $\ti h$ can  be easily eliminated, arriving at the final set of equations
\be \partial_+ A_-^i(g) -\partial_- A_+^i(g) -
\ti c_{kl}^{~~~i} A_-^k(g) A_+^l(g)=0,\ee
where $\ti c_{kl}^{~~~i}$ are the structure constants of the Lie algebra
$\cgt$. It can be directly checked, however, that the last equations are
just the field equations of the $\si$-model with the Lagrangian
\be L=E^{ij}(g)(g^{-1}\partial_- g)_i(g^{-1}\partial_+ g)_j.\ee

Equivalently, I may use the decomposition
\be l(\xi^+,\xi_-)=\ti g(\xi^+,\xi^-) h(\xi^+,\xi^-),\ee
where $\ti g\in \ti G$ and $ h\in  G$.
All  steps of the previous construction can be repeated,
to  end up with the dual $\si$-model
\be \ti L= \ti E_{ij}(\ti g)(\ti g^{-1}\partial_- \ti g)^i(\ti g^{-1}
\partial_+ \ti g)^j,\ee
where the matrix $\ti E_{ij} (\ti g)$ is defined as

\be \ti g^{-1}\ce^+ \ti g ={\rm Span}(\ti T_i + \ti E_{ij}(\ti g) T^j),
i=1,\dots,n,\ee
\be \ti g^{-1}\ce^- \ti g =
{\rm Span}(\ti T_i - \ti E_{ji}(\ti g) T^j), i=1,\dots,n.\ee
Before proceeding further,  note that the Poisson-Lie T-duality
is  a natural generalization of the
standard Abelian $R\to 1/R$ symmetry. Indeed, at the group origin ($g=e$
and $\ti g =\ti e$ ) the matrices $E(e)$ and $\ti E(\ti e)$ are
related
to each other as follows
\be E(e)\ti E(\ti e) = \ti E(\ti e) E(e)=1.\ee
The explicit dependence of $E$ on $g$ and $\ti E$ on $\ti g$ is given
by the matrices of the adjoint representation of $\cd$:
\begin{eqnarray} g^{-1}\ce^+ g={\rm Span}~ g^{-1}(T^i + E^{ij}(e)\ti T_j)g\cr
={\rm Span}\Big[(a(g)^i_{~l} +E^{ij}(e)b(g)_{jl})T^l +
E^{ij}(e)d(g)_j^{~l} \ti T_l\Big],\end{eqnarray}
where
\be g^{-1}T^i g\equiv a(g)^i_{~l} T^l, \qquad g^{-1}\ti T_j g \equiv
b(g)_{jl}T^l +d(g)_j^{~l} \ti T_l.\ee
Hence the $\si$-model matrix is given by
\be E(g)=(a(g) + E(e)b(g))^{-1}E(e)d(g).\ee
The matrix of the dual $\si$-model can be obtained in a completely
analogous way.

If there exists another maximally isotropic decomposition of the double $\cd$
in two subalgebras $\ck$ and $\ckt$ we obtain another pair of dual
$\si$-models with some matrices $E'(k)$ and $\ti E'(\ti k)$. This new dual
pair  again shares the equations of motions (4) with the old pair. Thus
we indeed see that the set of the mutually equivalent $\si$-models is
nothing but the set $M(\cd)$ introduced at the beginning of this section.

Note that the original field equations (4) have a huge symmetry
$l\to lm$ where $m$ is an arbitrary element of $D$. This symmetry is realized
in a {\it non-local} way in the $\si$-model language,
in particular the geometry
of the target does not possess it as an isometry. Only in the specific
cases of the Abelian and semi-Abelian duality is {\it part} of this symmetry
 realized locally and it does lead to the isometries of the target.
This is the reason why T-duality has so far  been considered to be connected
with the isometries of the target.

\section{\bf Poisson-Lie symmetry}

It is natural to pose a question when a $\si$-model with the free
$G$ action on its target admits the Poisson-Lie dual for some
dual group $\ti G$. As  was mentioned in the previous section, the
condition is the Poisson-Lie symmetry of the
$\si$-model,  defined as follows: Let
\be J_a=v_a^i(x)E_{ij}(x)\partial_+ x^j d\xi^+ -v_a^i(x)E_{ji}(x)\partial_-
 x^j d\xi^-\ee
be the `Noether' current 1-forms corresponding
to the right action of the group
$G$ on the target $\cm$ of the $\si$-model
\be S=\int d\xi^+ d\xi^-
E_{ij}(x)\partial_- x^i \partial_+ x^j .\ee
In these formulae
 $v_a^i(x)$ are the (left-invariant) vector fields corresponding to the
right action of $G$ on  $\cm$ and $x^i$ are some coordinates on $\cm$.
If the action of $G$ is isometry, then  the Noether currents
(22)  are closed 1-forms on the world-sheets of extremal strings.
If they are not closed but they rather obey on the extremal surfaces
a condition
\be  dJ_a={1\over 2}\ti c_a^{~kl}J_k \wedge J_l.\ee
(with $\ti c_a^{~kl}$ being the structure constants of some
Lie algebra $\cgt$),
we say that the $\si$-model has the $G$-Poisson-Lie symmetry with respect
to the group $\ti G$.
One can interpret the conditions  (24) also as (some of) the field
equations of the model, if the group $G$ acts transitively then they
exhaust all equations of motion.
Note also that the $\si$-models (13), defined in the previous section,
are $G$-Poisson-Lie symmetric with respect to the group $\ti G$, as  can
be directly seen from Eqs. (10)-(12).

The condition of the Poisson-Lie symmetry can be directly formulated
at the level of the $\si$-model Lagrangian. It reads \cite{KS2}
\be {\cal{L}}_{v_a}(E_{ij})=\ti c_a^{~kl} v_k^m v_l^n E_{mj}E_{in},\ee
where ${\cal L}_{v_a}$ stands for the Lie derivative corresponding
to the vector field $v_a$. The integrability condition for the Lie derivative
gives the `cocycle' condition of compatibility between the structure
constants of Lie algebras $\cg$ and $\cgt$ forming the double
\be \ti c_k^{~ac}c^l_{~fa}-\ti c_k^{~al}c^c_{~fa}-\ti c_f^{~ac}c^l_{~ka}
+\ti  c_f^{~al}c^c_{~ka}-\ti
c_a^{~lc}c^a_{~fk}=0.\ee
Needless to say, all  formulae (22)-(25) have their
dual counterparts, obtained just by exchanging the roles
of $\cg$ and $\cgt$.

Buscher's duality may be treated much in the same way
as  the atomic one; here
we give just the final form of the dual pair of $\si$-models.  The
 coordinates
labelling the orbits of $G$ in the target $\cm$, we denote
$y^{\al}(\al=1,\dots,k)$. The matrix of the $\si$-model $E_{ij}$ has
both
types of indices corresponding to $y^{\al}$ and $g$. The Lagrangian reads
\begin{eqnarray} &L=E_{\al\b}(y)\partial_- y^{\al}\partial_+ y^{\b}+
E_{\al b}(y,g)\partial_- y^{\al}(g^{-1}\partial_+ g)^b+\cr &+
E_{a\b}(y,g)(g^{-1}\partial_- g)^a\partial_+ y^{\b} +
E_{ab}(y,g)(g^{-1}\partial_- g)^a (g^{-1}\partial_+ g)^b.\end{eqnarray}
Note that the dependence of $E_{ij}$ on $g$ is fixed by  condition (25).
Explicitly
\be E(y,g)=(A(g)+E(y,e)B(g))^{-1}E(y,e)D(g),\ee
where $e$ is the unit element of $G$, $E(y,e)$ can
be chosen arbitrarily and
$A(g)$ is the $(k+dimG)\times(k+dimG)$ matrix
\be A(g)\equiv\left(\matrix{Id &0\cr 0&a(g)}\right),\qquad B(g)\equiv
\left(\matrix{0 &0\cr 0&b(g)}\right)
\ee
and $D(g)$ is given in terms of $d(g)$ in the same way as $A(g)$ in
terms of
$a(g)$. Of course, $a(g),b(g)$ and $d(g)$ are the same as in (20).
The $\si$-model matrix $E(y,g)$ given in (28) is in fact the most
general solution of the condition (25) in the adapted coordinates $(y,g)$.
As far as the dual model $\ti E$ is concerned:
\be \ti E(y,\ti g)=(\ti A(\ti g)+
\ti E(y,e)\ti B(\ti g))^{-1}\ti E
(y,e)\ti D(\ti g).\ee
Here
\be \ti E(y,\ti e)=(A+E(y,e)B)^{-1}(C+E(y,e)D),\ee
and
\be A=D=\left(\matrix{Id&0\cr0&0}\right),\qquad B=C=
\left(\matrix{0&0\cr0&Id}\right).\ee
\section{\bf Borelian double}
 The simplest non-Abelian double is the $GL(2,R)$ group; its Lie
algebra has the basis
\be T^1=\left(\matrix{1&0\cr 0&0}\right), \qquad
T^2=\left(\matrix{0&1\cr 0&0}\right),\ee
\be \ti T^1=\left(\matrix{0&0\cr 0&1}\right), \qquad
\ti T^2=\left(\matrix{0&0\cr -1&0}\right).\ee
Note that both sets of generators (33) and (34)  span
the Borelian subalgebras of the algebra $gl(2,R)$ (hence the name of the
double) and they are maximally isotropic with respect to the
non-degenerate
invariant symmetric bilinear form defined by the brackets
\be \la T^i,\ti T_j \ra=\delta^i_j, \qquad \la T^i, T^j \ra=
\la \ti T_i,\ti T_j \ra=0.\ee
We see that the Poisson-Lie duality will relate the $\si$-models with the
same target, namely the group manifold of the Borel group $B_2$ whose
Lie algebra is generated  by (33) or (34).
The following parametrization of the first copy (33) of $B_2$ is
convenient
\be g=\left(\matrix{e^{\chi}&\theta\cr 0&1}\right).\ee
The matrices $a(g),b(g)$ and $d(g)$ from Eq. (20) read
\be a(g)=\left(\matrix{1&e^{-\chi}\th\cr 0&e^{-\chi}}\right),\ee
\be b(g)=\left(\matrix{0&-e^{-\chi}\th\cr \th & e^{-\chi}\th^2}\right),
\ee
\be d(g)=\left(\matrix{1&0\cr -\th &e^{\chi}}\right),\ee
and the inverse $\si$-model matrix $E^{-1}(e)$ at the unit element of $B_2$ is
\be E^{-1}(e)=\left(\matrix{x&y\cr u&v}\right).\ee
By a direct application of  formula (21), the $\si$-model Lagrangian
(13) is worked out as follows:
$$ L={x\th^2+\th(u+y)+v\over \th^2+\th(u-y)+(xv-uy)}\partial_-\chi
 \partial_+\chi +{x\over \th^2+\th(u-y)+(xv-uy)}\partial_-\th
 \partial_+\th $$
$$- {y+(x-1)\th\over
\th^2+\th(u-y)+(xv-uy)}\partial_-\chi
 \partial_+\th - {u+(x+1)\th\over
\th^2+\th(u-y)+(xv-uy)}\partial_-\th
 \partial_+\chi.$$
With a slight abuse of the notation, the convenient parametrization
of the second copy (34) of $B_2$ is
\be \ti g=\left(\matrix{1&0\cr -\theta &e^{\chi}}\right)\ee
and the dual $\si$-model Lagrangian (15) has the same structure
$$\ti L=
{\ti x\th^2+\th(\ti u+\ti y)+\ti v\over \th^2+\th(\ti u-\ti y)+
(\ti x\ti v-\ti u\ti y)}\partial_-\chi
 \partial_+\chi +{\ti x\over \th^2+\th(\ti u-\ti y)+(\ti x\ti v-\ti u\ti y)}
\partial_-\th
 \partial_+\th $$
$$- {\ti y+(\ti x-1)\th\over
\th^2+\th(\ti u-\ti y)+(\ti x\ti v-\ti u\ti y)}\partial_-\chi
 \partial_+\th - {\ti u+(\ti x+1)\th\over
\th^2+\th(\ti u-\ti y)+(\ti x\ti v-\ti u\ti y)}\partial_-\th
 \partial_+\chi.$$
Here the parameters $\ti x,\ti y, \ti u$ and $ \ti v$ are related
to the original set $x,y,u$ and $v$ as
\be \left(\matrix{x&y\cr u&v}\right) \left(\matrix{\ti x&\ti y\cr\ti u&\ti v}
\right)= \left(\matrix{\ti x&\ti y\cr\ti u&\ti v}
\right)\left(\matrix{x&y\cr u&v}\right)=\left(\matrix{1&0\cr 0&1}
\right),\ee
which are nothing but  relations (18) in the context of this concrete
example.

\section{\bf Acknowledgement}
I thank A. Alekseev,
K. Gaw\c edzki, E. Kiritsis, S. Shatashvili, P. \v Severa and A. Tseytlin
for useful discussions.

\end{document}